\documentclass[prb,preprint,a4paper,showkeys,endfloats*]{revtex4}

\usepackage{amssymb}
\usepackage{graphicx}
\usepackage{mathptmx}

\begin{document}

\title{Ab initio study of magnetism at the TiO$_2$/LaAlO$_3$ interface}

\author{Mariana Weissmann}
\email{weissman@cnea.gov.ar}

\author{V. Ferrari}
\email{ferrari@tandar.cnea.gov.ar}

\affiliation{Departamento de F\'{\i}sica, Comisi\'on Nacional de Energ\'{\i}a
  Atomica, Avda. del Libertador 8250, 1429 Buenos Aires, Argentina}

\author{Andr\'es Sa\'ul}
\affiliation{Centre Interdisciplinaire de Nanoscience de Marseille, CNRS, Campus
  de Luminy, Case 913, 13288 Marseille Cedex 9, France}
\email{saul@ciman.univ-mrs.fr}

\begin{abstract}
  In this paper we study the possible relation between the electronic
  and magnetic structure of the TiO$_2$/LaAlO$_3$ interface and the
  unexpected magnetism found in undoped TiO$_2$ films grown on
  LaAlO$_3$. We concentrate on the role played by structural
  relaxation and interfacial oxygen vacancies.
  
  LaAlO$_3$ has a layered structure along the (001) direction with
  alternating LaO and AlO$_2$ planes, with nominal charges of +1 and
  -1, respectively. As a consequence of that, an oxygen deficient
  TiO$_2$ film with anatase structure
   will grow preferently on the AlO$_2$ surface layer. We
  have therefore performed ab-initio calculations for superlattices
  with TiO$_2$/AlO$_2$ interfaces with interfacial oxygen vacancies.
  Our main results are that vacancies lead to a change in the valence
  state of neighbour Ti atoms but not necessarily to a magnetic
  solution and that the appearance of magnetism depends also on structural
  details, such as second neighbor positions. These results are obtained
  using both the LSDA and LSDA+U approximations.

\end{abstract}


\date{\today }

\maketitle

\section{Introduction}

Thin films of transition metal oxides (ZnO, TiO$_2$, HfO$_2$) were
found to be ferromagnetic a few years ago, with a high Curie
temperature, when doped with magnetic ions
\cite{anatasaCo,zincCo,review}. The origin of this magnetic ordering
is still debated, but meanwhile ferro-magnetism was also found doping
with non magnetic ions, such as Cu \cite{Duhalde}, and also in undoped
cases \cite{hafnio,yoon,hassini}. Trying to understand this, several
experimental tests were performed in different laboratories: the
magnetic ion, the substrate over which the films were grown, the
substrate temperature and the oxygen pressure during film growth were
changed. The results are confusing, as some samples are magnetic while
others are not \cite{chambers}.  It now seems well established that
structural defects, such as oxygen vacancies, and low dimensionality
(thin films or nanoparticles) are necessary for the appearance of this
type of ferro-magnetism \cite{sudakar}.  One can understand why it is
difficult to obtain reproducible results from different experimental
samples, as it is hard to asses quantitatively the influence of
defects or disorder in any system.

At the same time, several ab-initio calculations have been performed
for these oxides in bulk \cite{nuestro1,nuestro2,pacchioni,kim}, with
impurities and/or defects, and different models have been proposed to
produce the spin ordering.  However, few calculations have discussed
if surfaces or interfaces have any influence in this type of magnetism
\cite{valeria}.

LaAlO$_3$ (LAO) single crystal is an excellent substrate for epitaxial growth of many oxides,
such as SrTiO$_3$, BaTiO$_3$, PbTiO$_3$ and TiO$_2$. While the surface electronic structure and
atomic relaxation of those oxides \cite{eglitis} and the 
SrTiO$_3$/LAO  interface \cite{freeman,pickett} have been widely studied, the TiO$_2$/LAO
interface has not. In the present paper we focus on undoped films of TiO$_2$ grown on top
of a LAO substrate. This system has been found to be
ferromagnetic, with a Curie temperature of about 800K
\cite{yoon,hassini}, grown either by pulsed laser deposition or by
spin coating. It is also well suited for ab-initio calculations
because LAO in its cubic phase and TiO$_2$ in anatase structure, both along the
(001) direction, have a very small structural misfit. In fact, thin
films of TiO$_2$ using LAO as a substrate develop anatase structure,
indicating epitaxial growth.

The aim of this work is to study the role played by interfacial oxygen
vacancies on the magnetic properties of the interface. For this
purpose, we consider a model system with a high concentration of
vacancies. Under certain conditions we find a localized magnetic moment in the
interfacial Ti atoms, depending not only on the presence of vacancies,
but also on subtle structural details.  We do not consider here the
possible interaction between these localized moments.

\section{The system}

To simulate interfaces we use a super-lattice geometry, so that the
system under study results periodic in 3D : ... LAO / TiO$_2$ / LAO /
TiO$_2$ / ...

\begin{figure}[htb]
\begin{center}
  \includegraphics[width=0.65\textwidth]{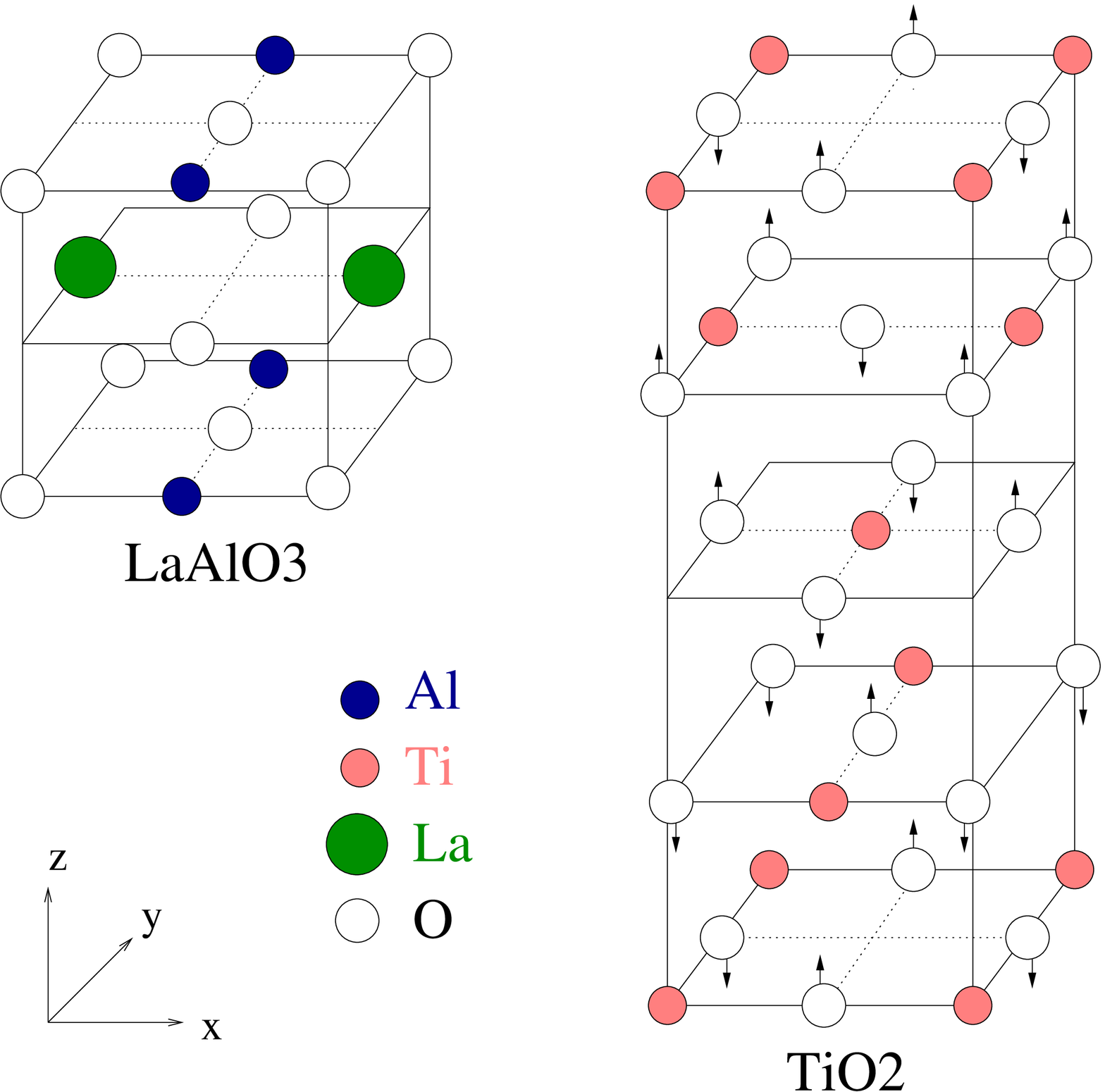}
\caption[]{\label{fig:1} 
  Unit cells of the two component materials in bulk: LAO (left) and
  TiO$_2$ anatase (right). The arrows on the oxygen atoms of anatase
  indicate the off-plane displacement with respect to the Ti plane.}
\end{center}
\end{figure}
Figure \ref{fig:1} shows the unit cells of the two component
materials: LAO and TiO$_2$ anatase. LAO is a layered material and
along the (001) direction. The planes LaO and AlO$_2$ have nominal
charges of +1 and -1, respectively. TiO$_2$ anatase is almost layered:
the oxygen atoms are slightly out of the Ti planes, as indicated by
the small arrows in Fig.~\ref{fig:1}. Each plane has zero nominal
charge, so that the interface with LAO will restructure due to
electrostatics either changing the valence of Ti from Ti$^{4+}$ to
Ti$^{3+}$ or Ti$^{2+}$ or by missing oxygen atoms. In this respect,
our previous preliminary work \cite{valeria} showed, using total
energy calculations, that it is easier to remove an oxygen atom from
the anatase side of the interface than from the LAO side.

\begin{figure}[htb]
\begin{center}
  \includegraphics[width=\textwidth]{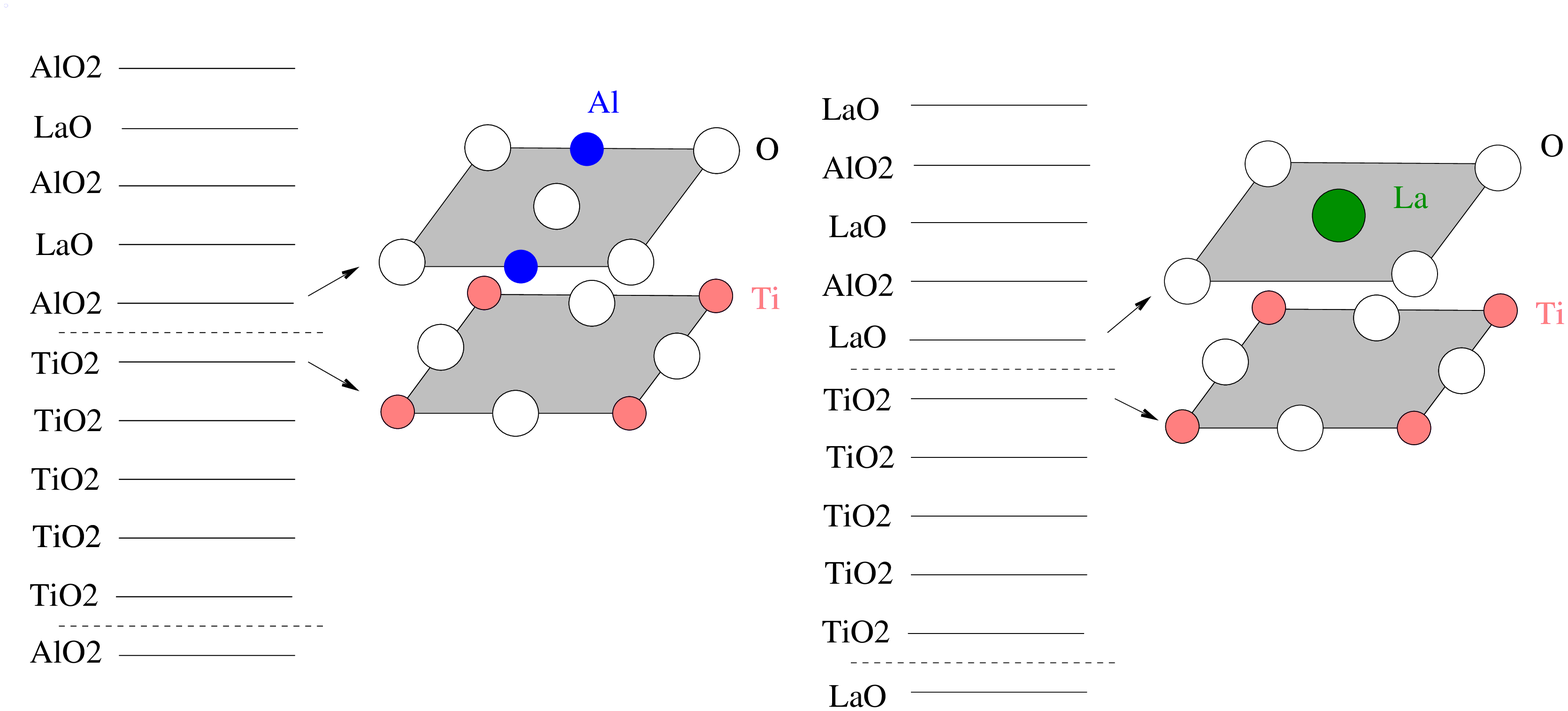}
\caption[]{\label{fig:2} Schematic diagram of the two possible 
  ways of stacking to form a super-lattice: AlO$_2$ facing TiO$_2$
  (left) and LaO facing TiO$_2$ (right). The dotted lines indicate the
  position of the interfaces.}
\end{center}
\end{figure}
\begin{figure}[htb]
\begin{center}
  \includegraphics[width=0.7\textwidth]{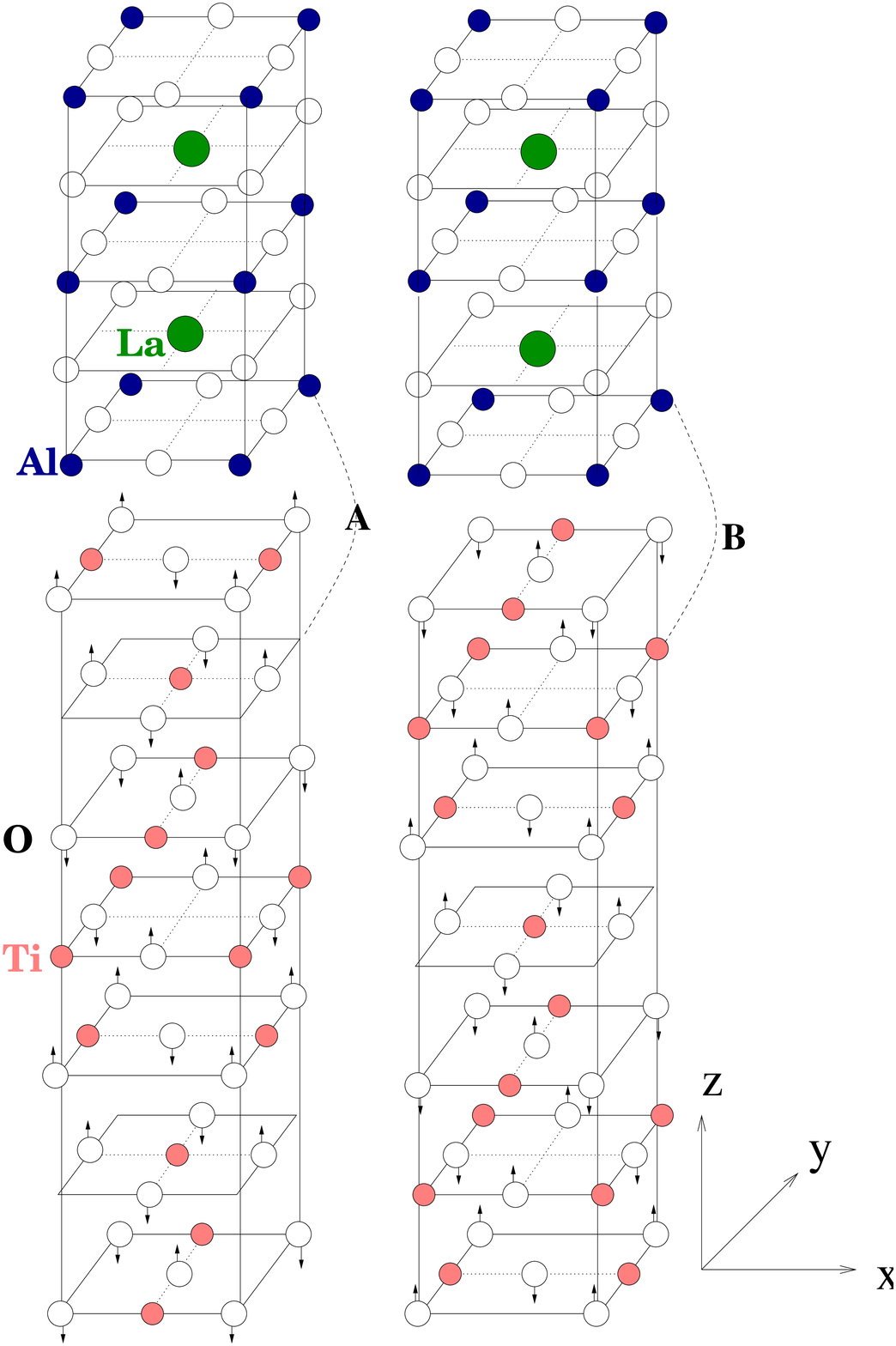} 
\caption[]{\label{fig:3} Unit cells of the two super-lattices considered
  in this work. Due to the off-plane oxygens in the anatase structure
  and also to the differences in the second neighbor structure (shown
  by dotted lines) there are two types of AlO$_2$/TiO$_2$ interfaces,
  A (left) and B (right).}
\end{center}
\end{figure}

Figure \ref{fig:2} shows the two possible interfaces, namely the two
ways of stacking to form the super-lattice, that maintain the
octahedral environment of the Ti atoms.  Without oxygen vacancies, our
previous calculations \cite{valeria} showed that neither one has a
magnetic solution. With oxygen vacancies near the interface it is
clear, due to electrostatics, that the AlO$_2$/TiO$_2$ interface will
be preferred with respect to the LaO/TiO$_2$ one. While there is only
one way of stacking in which LaO faces TiO$_2$, there are two ways
with AlO$_2$ facing TiO$_2$, that we name A and B. They are shown in
Fig.~\ref{fig:3}, that presents the super-lattice unit cells used in
the present calculations.  In order to have all interfaces in the
super-lattice either of type A or type B, 7 anatase planes and 5 LAO
planes are considered. If there are no oxygen vacancies, type A is
clearly more stable, with lower total energy, as Al-O bonds are formed
at the interface due to the off plane oxygens in the anatase
structure. With oxygen vacancies, each distribution has to be studied
separately.

The size of the super-cells is obtained by minimizing the total energy
in each case, and the atomic positions inside the cell are allowed to
relax but only in the direction of the super-lattice. The reason for
this last restriction is based on our previous results, apart from making the calculation more feasible.  In fact, the calculations on pure bulk rutile
TiO$_2$ \cite{nuestro2} showed that oxygen vacancies can give rise to
a local magnetic moment due to structural relaxation. However, this is
only true for a fairly large concentration of vacancies and for a
fixed cell size, which means there must be some restriction to the
full structural relaxation. We could not find a similar result for
bulk anatase TiO$_2$, but our assumption here is that the LAO
substrate could impose such a structural restriction to the anatase
film thus giving rise to magnetic solutions. Therefore, the
calculations in the present paper are performed fixing the unit cell
size in the direction normal to the stacking to that of cubic LAO
(lattice constant = 3.8 \AA) and relaxation is only allowed in the
direction of the super-lattice.

\section{Method of Calculation and Results}

For the Density Functional Theory (DFT)~\cite{DFT} calculations we have used the Wien2k code,~\cite{wien2k} that is an implementation of the FPLAPW (full potential linear augmented plane
waves) method, in which the space is
divided into muffin tin  spheres around the atoms and an interstitial region. Plane waves
are used to describe the region outside the spheres. The number of plane waves in the interstitial region is set by the cut-off  parameter RK$_{max}$. In this work we use RK$_{max}$=7 that corresponds to an energy cutoff of 340 eV.
We consider both the local spin density approximation
LSDA \cite{LDA} and the LSDA+U \cite{LDAplusU} one.  The calculation is scalar relativistic and the atomic
sphere radii  are taken to be small, so that the atomic
spheres do not overlap when relaxing the structure. They are 1.7 a.u.
for Ti and Al, 1.4 a.u. for O and 2.5 a.u. for La. The number of
k-points in the Brillouin zone is 50 or 100 for the relaxation
procedure and increased to 200 k-points for the relaxed structure.

\begin{figure}[htb]
\begin{center}
  \includegraphics[width=0.7\textwidth]{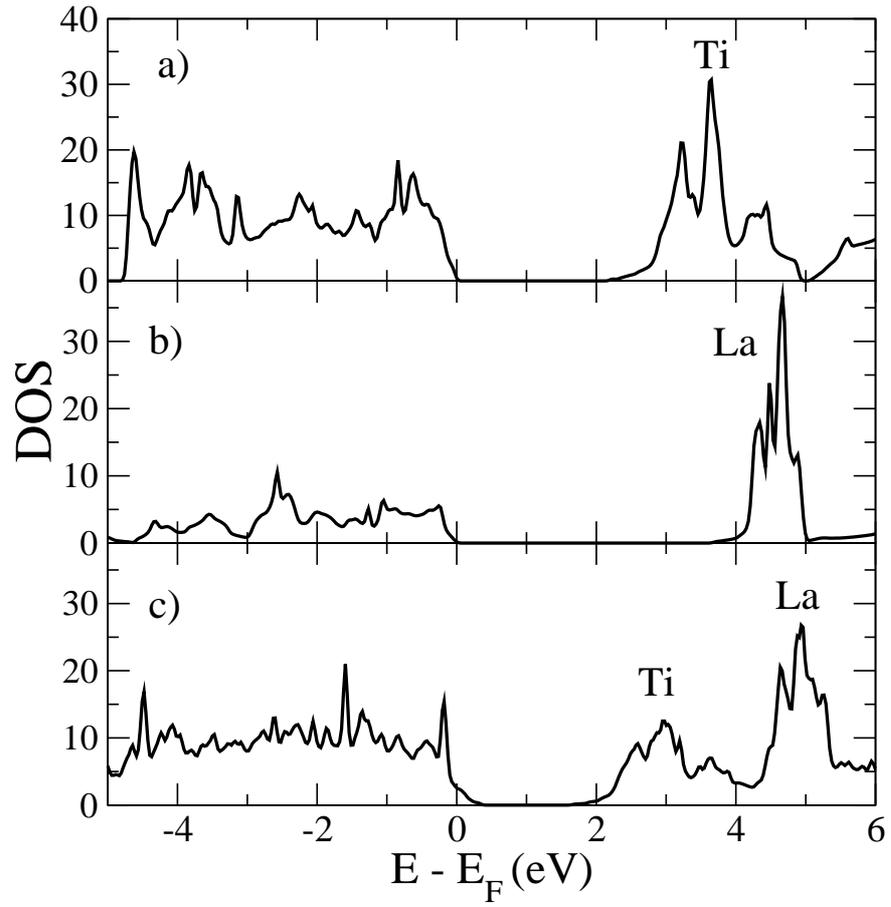} 
\caption[]{\label{fig:4} Calculated densities of states of: 
  (a) bulk TiO$_2$ anatase 
  (b) bulk LAO,
  (c) Super-lattice with AlO$_2$/TiO$_2$ interfaces (relaxed structure).}
\end{center}
\end{figure}
Figure \ref{fig:4} shows the densities of states calculated within
LSDA for the component systems and for the super-lattice. There are no
magnetic solutions for any of these cases. The gap
obtained for anatase is 2.2 eV, to be compared with the experimental
3.2 eV, the gap for LAO is 4.3 eV to be compared with the 5.6 eV
experimental value. This feature of the LSDA approximation, that gives
small band gaps, can be corrected using the LSDA+U method or hybrid functionals \cite{Pacchioni}. 
 The LSDA+U approach introduces an
additional term based on a simple Hubbard model for electron
on-site repulsion. This removes the self-interaction
error by energetically penalising partial occupation of
the relevant electronic states, at the price of introducing an empirical parameter in an ab-initio calculation.

\begin{figure}[htb]
\begin{center}
  \includegraphics[width=0.7\textwidth]{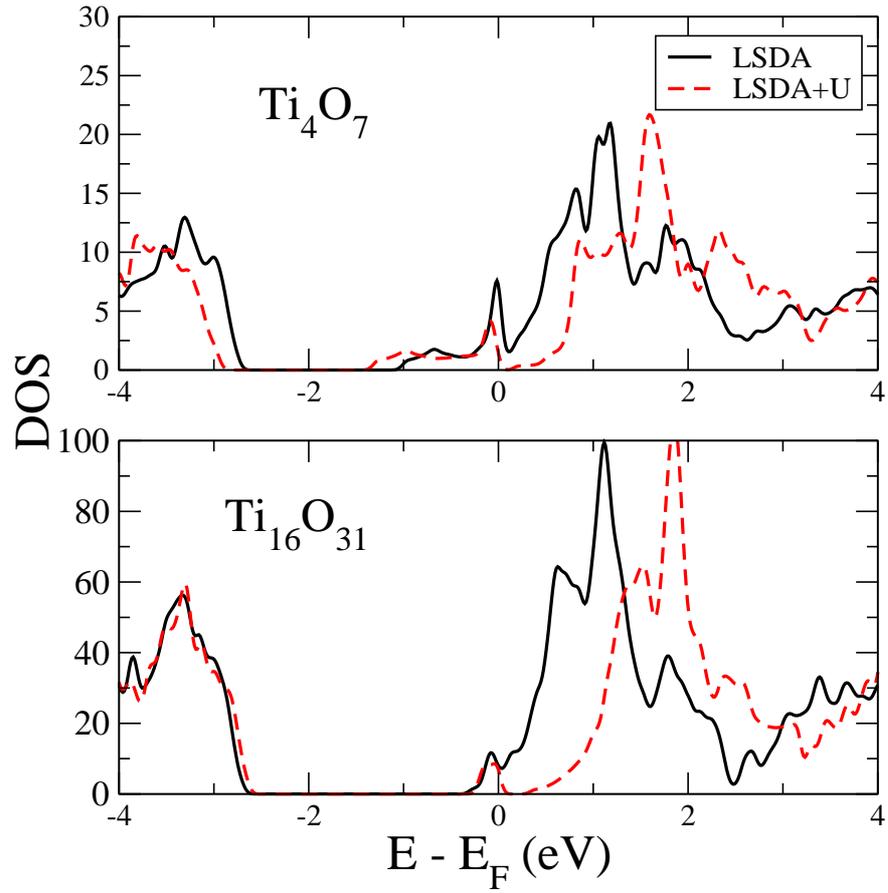}
\caption[]{\label{fig:5} Calculated densities of states of bulk 
  anatase, with two different concentrations of oxygen vacancies in
  the LSDA and LSDA+U approximations (U = 5 eV). The structure is
  relaxed separately in each case.}
\end{center}
\end{figure}

Before studying the effect of oxygen vacancies at the interface, we
start with bulk TiO$_2$.  Figure \ref{fig:5} shows the densities of
states of bulk TiO$_2$ anatase with one oxygen vacancy in the unit
cell, for two different concentrations and both methods of
calculation, relaxing all structures separately. 
No magnetic solutions are found for any of these cases.
A defect state
appears inside the band gap, too close to the conduction band in the
LSDA approximation, thus making the system metallic. In the LSDA+U
approximation, using U = 5 eV, the band gap  
increases by about 1eV and the system becomes a semiconductor, as the defect state
is separated from the conduction band.  However, using only one
parameter U (for the $d$ electrons of Ti) it is not possible to fit both the
experimental  band gap and the position of the defect state
inside the gap (which is at 0.8 eV from the conduction band). It has been shown for reduced rutile TiO$_2$ surfaces that a discontinuity appears for this
property as a function of U at 4 eV and that a larger value is required to ensure the semiconductor character of the system as well as the correct position of the gap state \cite{morgan}.
At this point it is important to remark that in the present work we consider a large concentration of 
vacancies in the interfaces, with   the defect
state developing into an impurity band, and therefore the exact position of the defect state is not so relevant.

\begin{figure*}[htb]
\begin{center}
  \includegraphics[width=1\textwidth]{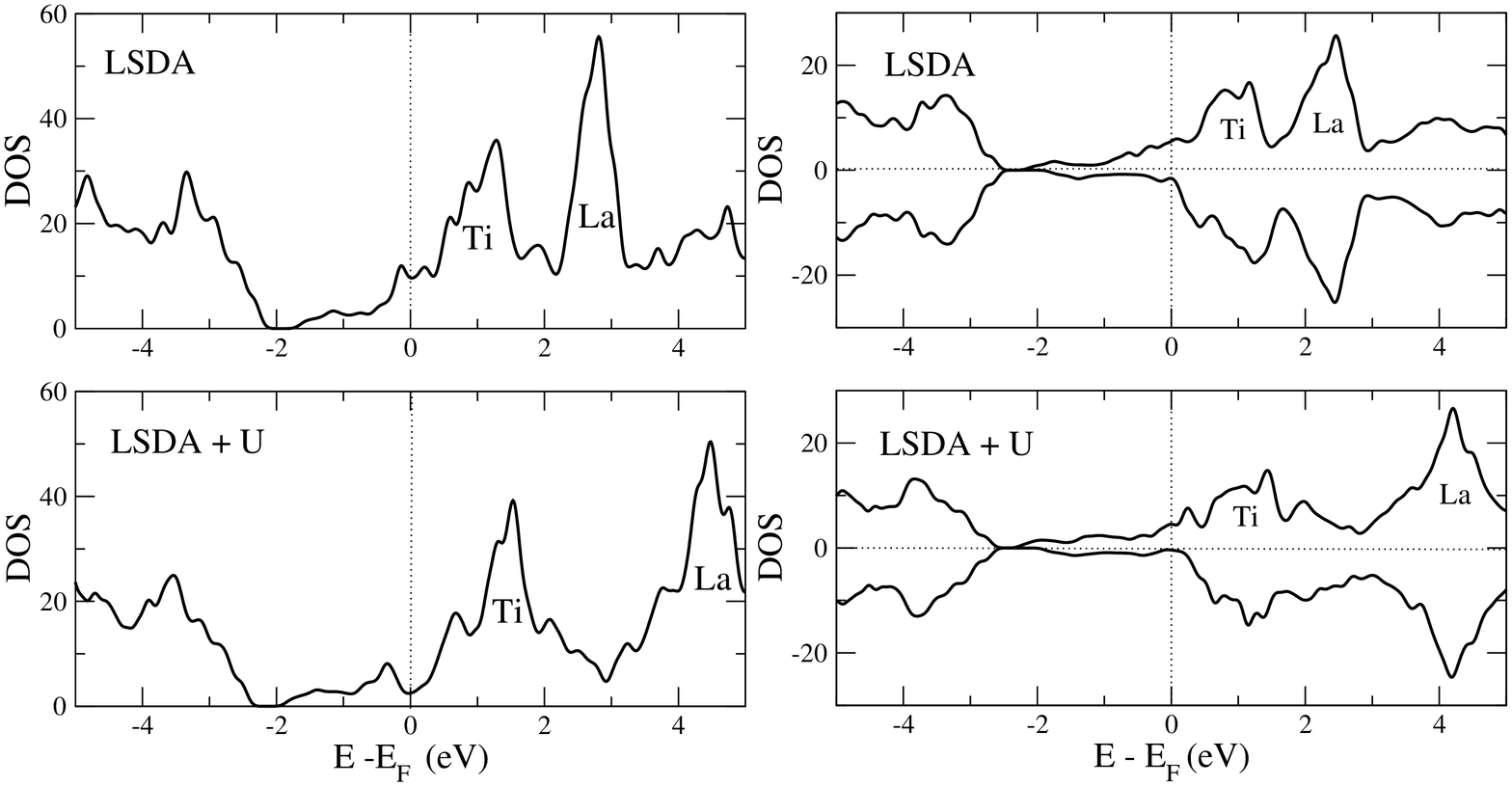}
\caption[]{\label{fig:6} Calculated densities of states of 
  super-lattices A (left) and B (right) in Fig.~\ref{fig:3}. with no
  oxygen atoms in the Ti side of the interface, using the LSDA (top)
  and LSDA+U (bottom) approximations (U = 5 eV). Case A is not
  magnetic, so that the contributions of both spin types have been
  added. In case B we show spin up and spin down contributions
  separately, the magnetic moment of the unit cell is 2.7 $\mu_B$ and
  inside the muffin tin of each inter-facial Ti it is 0.65 $\mu_B$.}
\end{center}
\end{figure*}

Turning now to vacancies at the interface, Fig.~\ref{fig:6} shows the
densities of states for the super-cells A and B of Fig.~\ref{fig:3}
when all oxygen atoms on the anatase side of the interface have been
removed, calculated with LSDA and with LSDA+U.  This extreme case is
very interesting, as the total energy of systems A and B are equal within the
calculation error (1 mRy) but system B is magnetic while system A is
not. The only difference between the two structures lies on the second
neighbor layer of the interface, as shown by a dotted line in
Fig.~\ref{fig:3}. The local distortion is similar in both cases,
producing a ripple in the inter-facial AlO$_2$ layer.

\begin{figure}[htb]
\begin{center}
  \includegraphics[width=0.7\textwidth]{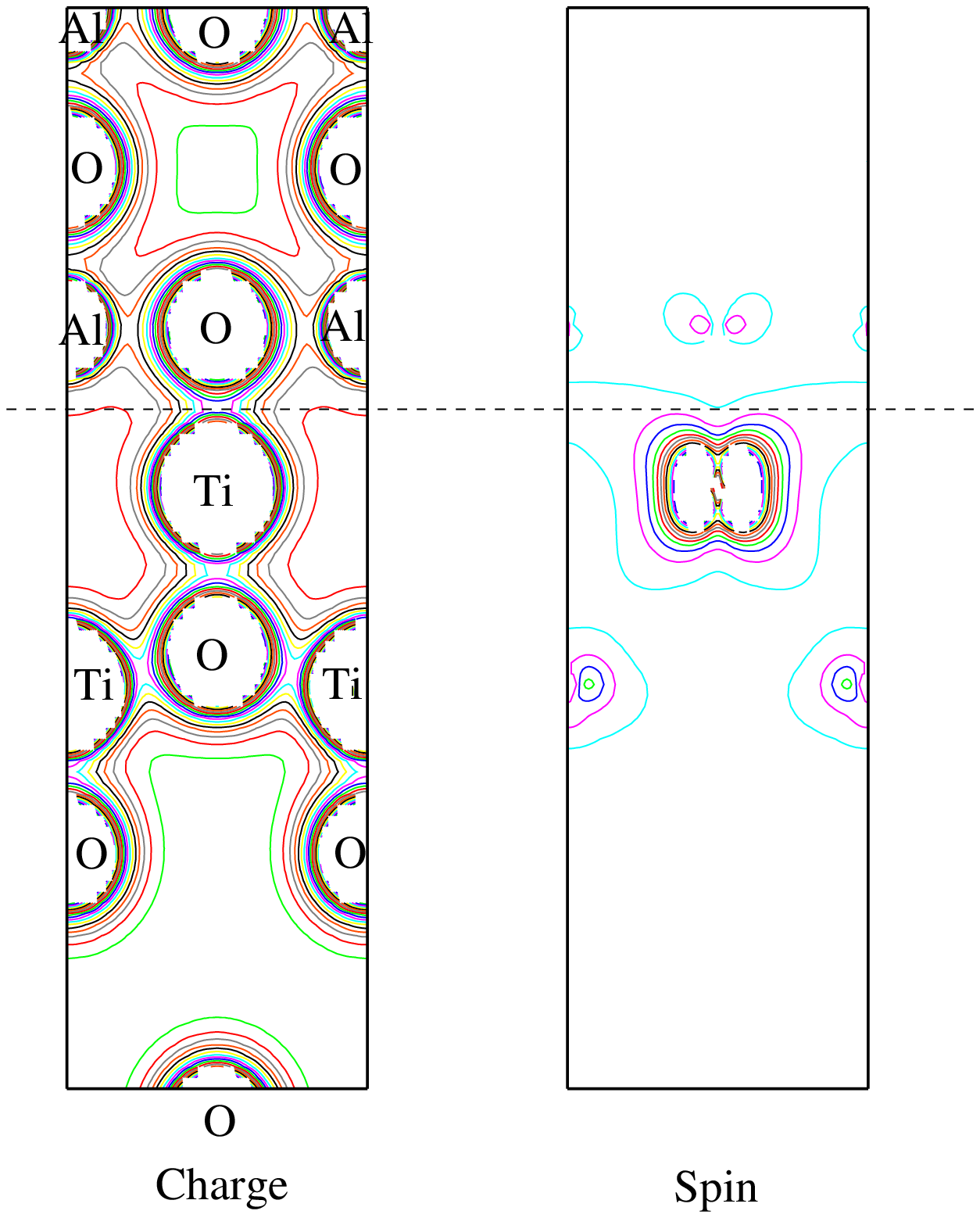}
\caption[]{\label{fig:7}
  Charge and spin density maps for super-lattice B, near the
  AlO$_2$/TiO$_2$ interface with 2 oxygen vacancies. The effect of
  relaxation is too small to be seen in this scale, but the AlO$_2$
  plane is buckled. Magnetism is mostly localized at the inter-facial
  Ti atom, spin isolines in the figure start at 0.01 and are spaced by
  0.03. Charge isolines start at 0.1 and are spaced 0.1, The interface
  is indicated by a dotted horizontal line.}
\end{center}
\end{figure}
The magnetic moment in case B is mostly located at the inter-facial
Ti, as can be seen in Fig.~\ref{fig:7}. However, the total occupation
of the $d$ orbitals in this Ti atom is very similar in cases A and B,
both can be considered Ti$^{3+}$, calculated with LSDA and LSDA+U.  In
both approximations, the defect level widens to form a band that
almost fills the band gap completely, due to the large number of
oxygen vacancies.  This wide feature inside the band gap is also found
in XPS experiments \cite{rumaiz}.

\section{Discussion and Conclusions}

The results shown in the previous section call our attention to the
following fact: oxygen vacancies lead to a change of valence in a
nearby Ti atom turning it into Ti$^{3+}$, but this does not
necessarily imply that the system will have a magnetic solution.
Trying to understand this unexpected result, we resort to the well
known Stoner criterion.  The main idea behind it is that hybridization
and band splitting are competing mechanisms to lower the energy of the
system. For this purpose we performed non-magnetic calculations of
super-lattices A and B, in the unrelaxed structures.

\begin{figure}[htb]
\begin{center}
\includegraphics[width=\textwidth]{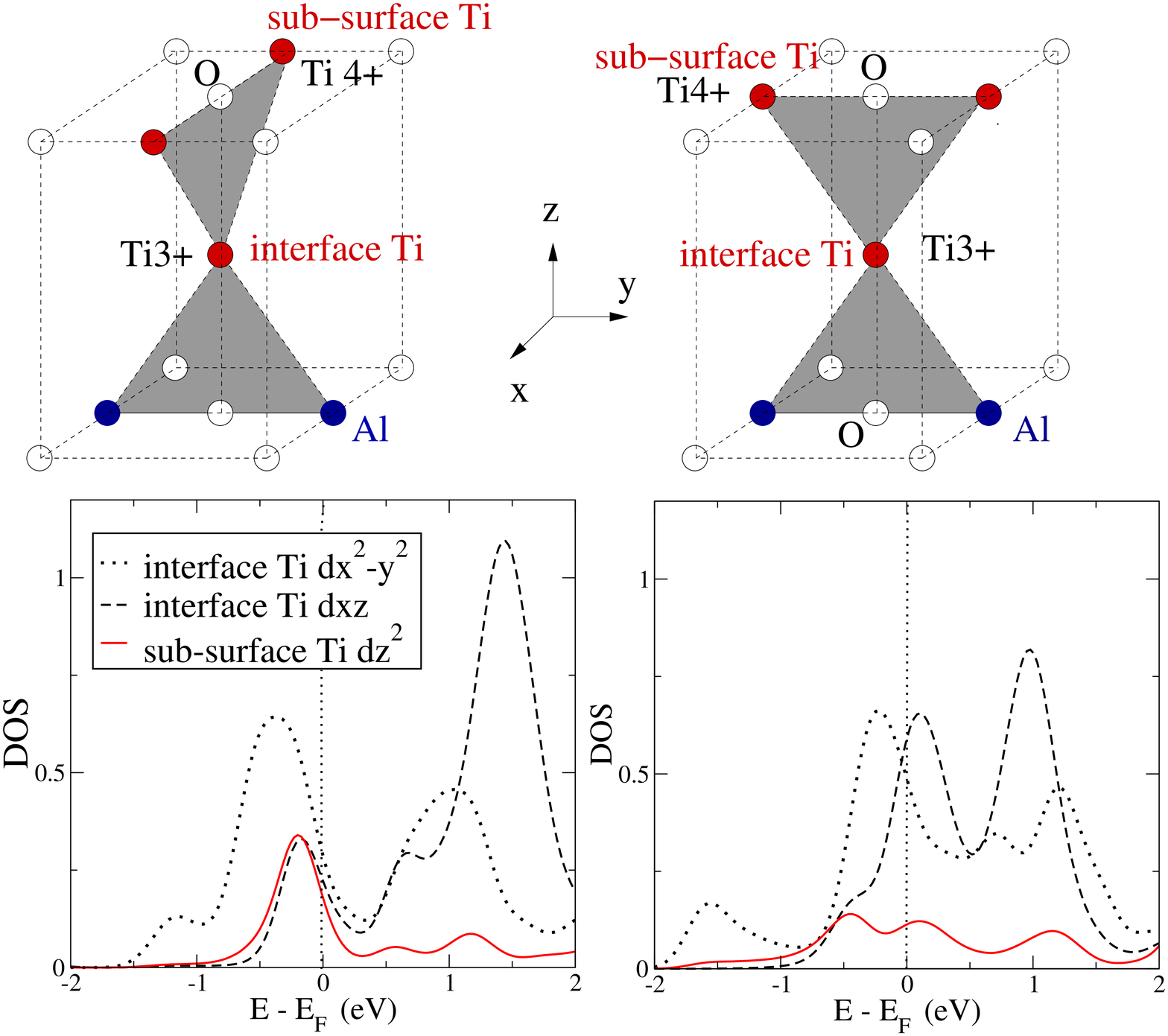} 
\caption[]{\label{fig:8} Non-magnetic calculations for
  super-lattices A (left) and B (right). Top: atomic structures in the
  proximity of the interfaces. Bottom: local densities of states for
  an energy range close to the Fermi energy. Dotted line:
  $d_{x^2-y^2}$ of the inter-facial Ti, broken line: $d_{xz}$ of the
  inter-facial Ti and full line: d$_{z^2}$ of the subsurface Ti atom.
  Interface A shows a much stronger hybridization (bonding) between
  orbitals of inter-facial and subsurface Ti atoms.}
\end{center}
\end{figure}
\begin{figure}[htb]
\begin{center}
\includegraphics[width=\textwidth]{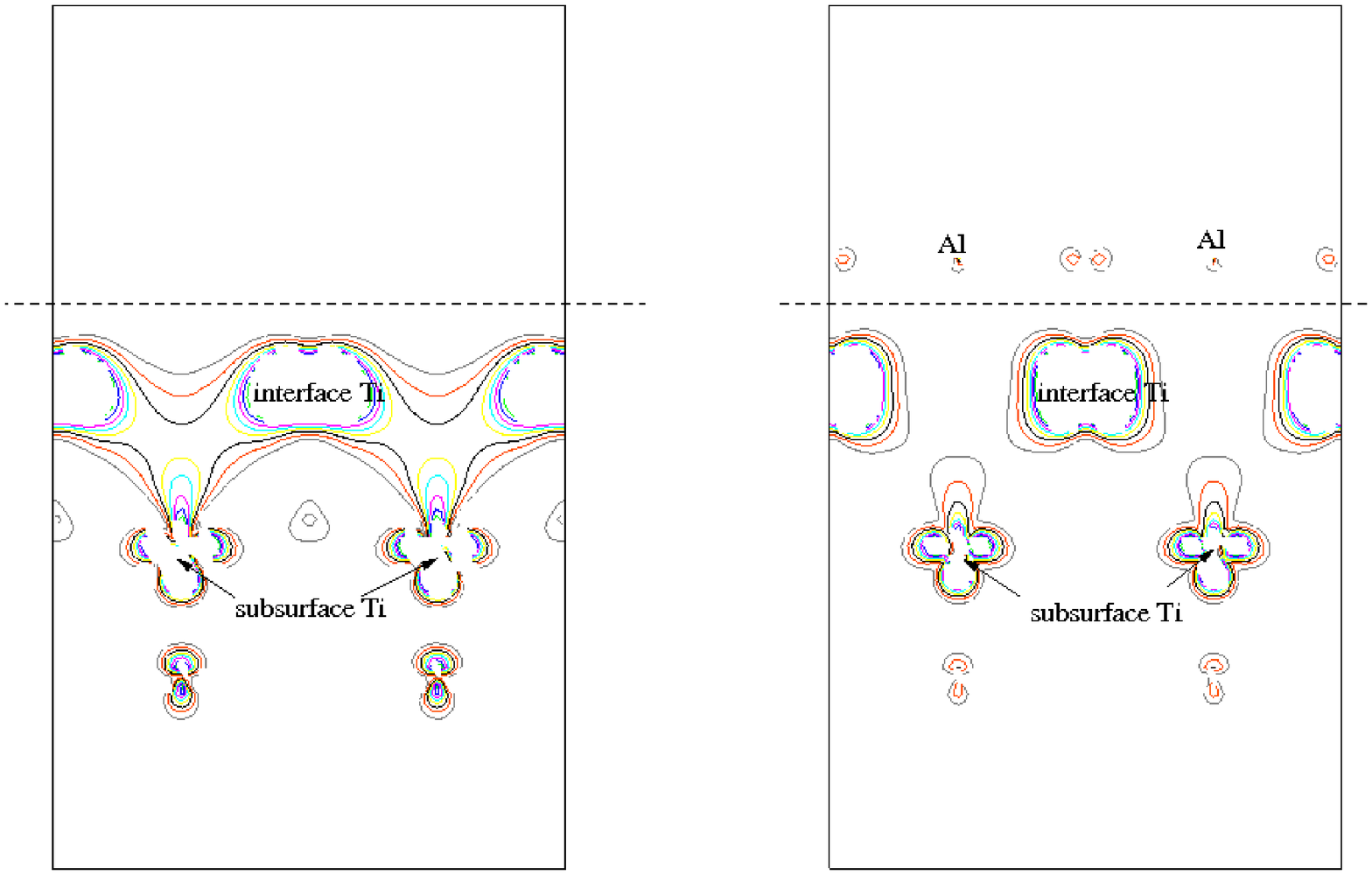}
\caption[]{\label{fig:9}  Charge densities at interfaces A (left) and
  B (right) in the energy range $E_f-0.5$eV$<E <E_f$, plotted for
  $yz$-plane of Fig.~\ref{fig:8}. The isolines start at 0.04 and are
  spaced by 0.02.}
\end{center}
\end{figure}
\begin{figure}[htb]
\begin{center}
  \includegraphics[width=0.70\textwidth]{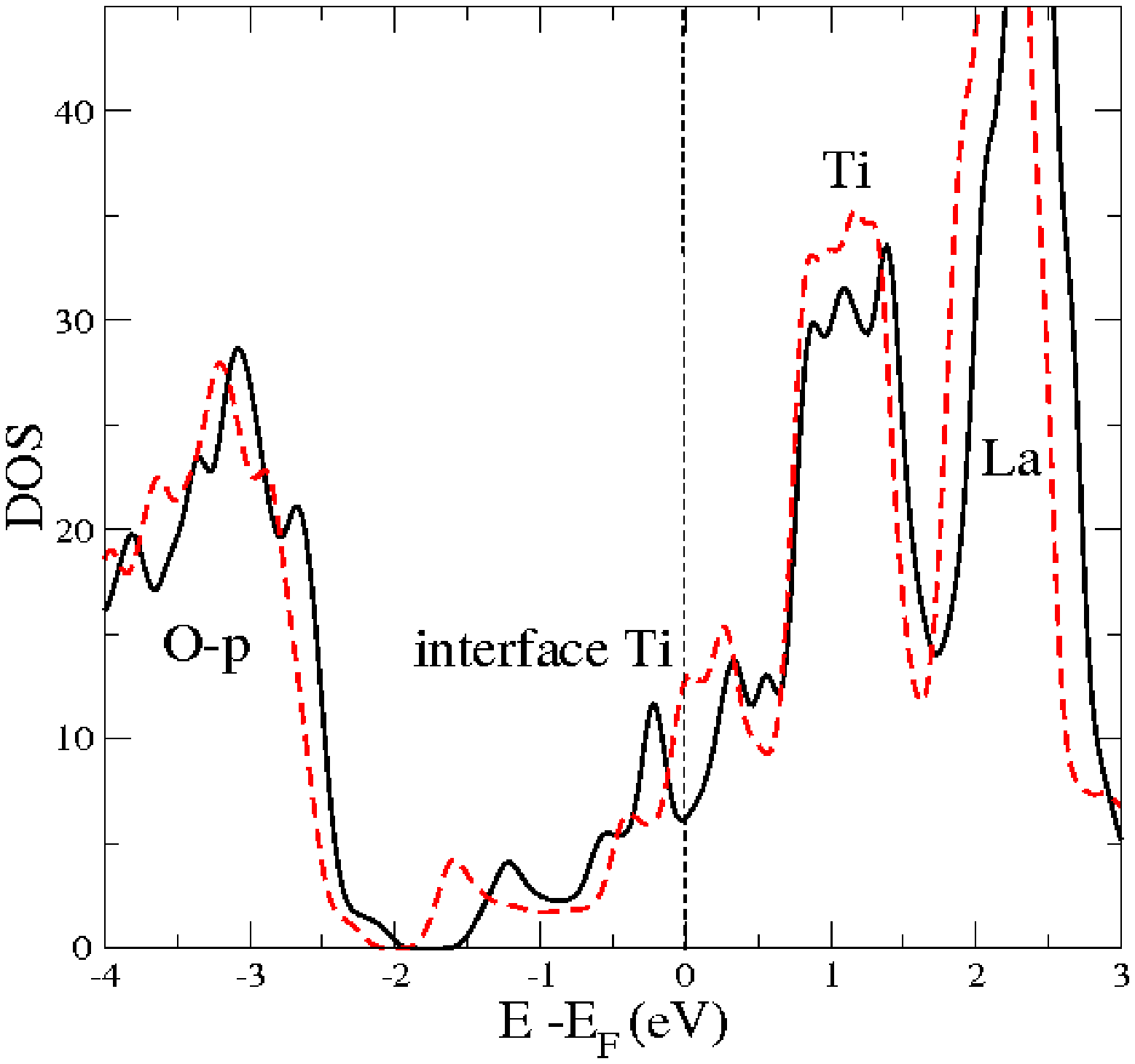} 
\caption[]{\label{fig:10} Calculated non magnetic total densities of 
  states for super-lattices A (full line) and B (dotted line). At the
  Fermi level their values are 7.2 for case A and 11.5 for case B,
  indicating that case B has a larger probability of being magnetic.}
\end{center}
\end{figure}

The results are shown in Figures \ref{fig:8}, \ref{fig:9} and
\ref{fig:10}. The stronger hybridization in case A is quite obvious in
Fig.~\ref{fig:8}: the partial density of states of orbital d$_{xz}$ of
the inter-facial Ti overlaps almost completely with that corresponding
to orbital d$_{z^2}$ of the subsurface Ti. In case B, the Al atom that
is in the same position as the Ti subsurface atom, destroys this
hybridization. It is the geometric difference in second neighbor
positions which drives the system to a magnetic solution.

Another way of looking at this is by plotting the charge density, in an energy range close to the Fermi energy. This is shown in  Fig.~\ref{fig:9}, and as expected it is
quite different for both structures.

Fig.~\ref{fig:10} shows that the non-magnetic total density of states
at the Fermi level is larger in case B than in case A, in agreement
with the presence of magnetism.

The study of this extreme example, with such a large number of
inter-facial defects, shows that oxygen vacancies and consequently the
presence of Ti$^{3+}$, are a necessary but not sufficient condition to
obtain magnetic solutions.  This result may be related to the
difficulty of producing reproducible experimental results.

\section*{Acknowledgements}

We acknowledge support by CONICET under grant PIP 112-200801-38,
ANPCyT under grant PICT-06-157 and a CNRS-CONICET International
Scientific Cooperation Project (PICS).

\end{document}